\documentclass[twocolumn,showpacs,prl,aps]{revtex4}
\usepackage{graphicx}

\newcommand{\be}{\begin{equation}}
\newcommand{\ee}{\end{equation}}
\newcommand{\ba}{\begin{eqnarray}}
\newcommand{\ea}{\end{eqnarray}}
\newcommand{\ban}{\begin{eqnarray*}}
\newcommand{\ean}{\end{eqnarray*}}

\newcommand{\braket}[2]{\mbox{$ \langle #1 | #2 \rangle $}}

\newcommand{\ket}[1]{\mbox{$ | #1 \rangle $}}
\newcommand{\bra}[1]{\mbox{$ \langle #1 | $}}

\begin{document}

\title{\bigskip Comments on the experimental disproof of Multisimultaneity}

\author{ Olivier Costa de Beauregard}

\affiliation{\it Fondation Louis de Broglie, 23 Rue Marsoulan
75012 Paris}

\date{December 23, 2001}

\begin{abstract}
The recent Geneva experiment strikingly displays \emph{lawlike
reversibility} together with \emph{quantum nonseparability}. As in
any probabilistic physics \emph{correlation} expresses
\emph{interaction}, and as Born's probability rules grafted upon
de Broglie's wave mechanics turn the probability scheme into the
code of an information transmitting telegraph, the Lorentz and CPT
invariant transition amplitude reversibly carries a zigzagging
causation.
\end{abstract}


\maketitle

Stefanov's et al. \cite{[1]} valuable experiment exemplifies once
again the counterintuitive phenomenology of non-separability
inherent in the Born-Jordan wavelike probability calculus; it
stems from the cross, interference, terms present in probability
expressed as absolute-squared amplitude. Algebraic nonseparability
entails geometric nonlocality; emphasis on its time aspect can be
worded atemporality.

Lawlike reversibility versus factlike irreversibility \cite{[2]}
of cause, that is, elementary level law versus macroscopic fact,
is crucial. The latter is mere jurisprudence ; to turn
juris-prudence into law is a fatal mistake. Euler clarified
mechanical law in terms of extremed action, and Bayes
probabilistic law via reversal of conditionals.

Mechanical-and-probabilistic reversibility survives the
Born-Jordan revolution ; it is expressed via the Hermitian
reversibility $\braket{\varphi}{\psi}= \braket{\psi}{\varphi}^*$
of a transition amplitude. In an $(x,ct)$ picture the
$\varphi\leftrightarrow\psi$ exchange renders PT reversal, and
conjugation the C exchange; so Hermitian reversibility renders CPT
reversibility.

Quantum transitions are between representations \cite{[5]}, which
are not ``realistic'' but ``Picasso style'' due to
complementarity. Prepared $\bra{\varphi}$ and measured
$\ket{\psi}$ (retropared says Hoekzema \cite{[6]} representations
are correlated by an N-uple transition amplitude pictured as a
Feynman graph; turning upside down Stefanov's et alii wording I
state: ``Lorentz-and-CPT invariance plus topological invariance of
a correlation amplitude are primary traits inherent in the
reversible causality concept making no difference between cause
and effect''.

For example a photon transiting between two linear polarizers is
neither in the prepared $\bra{\varphi}$ nor in the measured
$\ket{\psi}$ representation, but is reversibly transiting between
them; a time extended interference exists between  $\bra{\varphi}$
and $\ket{\psi}$. The Geneva experiment \cite{[1]} displays such a
spacetime extended interferometry over a macroscopic interval.

The wavelike correlation between prepared (emitted, coded) and
retropared (received, decoded) representations is reversibly
carried as a signal via the network pictured as a Feynman's graph.
A concise derivation of the EPRB correlation formula (direct,
inverse, or space-time transposed) is thus possible \cite{[7]}.

Reciprocity of the twin faces, cognizance and organization, of
information shows up in the reciprocal interventions of pre- and
retro-paration; a questions-and-answers game between reality and
representation is thus going on, where coding impresses
organization and decoding expresses knowledge (reality precedes in
decoding, realization follows in coding).

\emph{Lawlike reversibility-but-factlike irreversibility}
\cite{[2]} is formalized as information-negentropy equivalence
$N/I = k \log 2$, with $I$ expressed in bits and $N$ in
``practical'' thermal units (say clausius); \emph{law is expressed
by the finiteness, fact by the smallness of $k$}. $\mathcal{N}$
denoting Avogadro's number and $R$ the constant in the law of
perfect gases $pv = RT$, one has $k= R/\mathcal{N}$ (incidentally
one wonders why the classics did not set $R=1$ by definition of
temperature read on a perfect gas thermometer). Anyhow $k$ is very
small ``because $\mathcal{N}$ is very big''; so knowledge is
extremely cheap and organization expensive; said otherwise :
``cognizance is normal and psychokinesis \cite{[8],[9]}
paranormal''. If $k$ were zero cognizance would be cost-free and
free-will an illusion -a hypothesis that was implicit in the
theory named ``epiphenomenal consciousness''. Interestingly,
Born's wavelike probability scheme associates \cite{[10],[11]}
retarded or advanced causation with statistical pre- or
retro-diction.

Postulating, as previously said, that probability together with
information its alter ego are psycho-physical invalidates Jaynes
\cite{[9]} severance of ``ontology and epistemology''; it
identifies joint probability to reversible interaction. The
quantum paradigm then likens the wavelike probability calculus to
the code of an information transmitting telegraph.

I repeatedly said to Suarez that I could not buy his
``multisimultaneity'' hypothesis which, in his own \cite{[1]}
wording, is ``conceptually foreign to both quantum mechanics and
relativity''. So, once again a definite quantum prediction looked
so incredible that an ad hoc counter proposal was elaborated,
tested, and refuted, which of course is highly significant
\cite{[13]}.

Decoherence at reception (decoding, measurement) and the one
imposed at emission (coding, preparation) is a calculation
approximation yielding the specific ``empirical reality''
\cite{[12]} selected by the experimental setup. A more subtle
essence thus thrown out is the one reversibly telegraphed between
preparing and measuring physicists.

Paraphrasing the cosigners \cite{[1]} I deem \emph{quantum
correlation to be a basic concept synonymous to back and forth
telegraphed information} (knowledge-and-organization) that is,
\emph{to zigzagging causation.}

\end{document}